\documentstyle[11pt,fleqn,epsf]{article}
\def\nat#1#2#3{     {\it Nature }{\bf{#1},} (#2) #3 }
\def\ast#1#2#3{     {\it Astron. J }{\bf{#1},} (#2) #3 }
\def\astp#1#2#3{     {\it Astrophys. J }{\bf{#1},} (#2) #3 }
\def\aastp#1#2#3{     {\it A. Astrophys. J }{\bf{#1},} (#2) #3 }
\def\cqg#1#2#3{     {\it Class. Quantum Grav. }{\bf{#1},} (#2) #3 }
\def\grg#1#2#3{     {\it Gen. Relativ. Gravit. }{\bf{#1},} (#2) #3 }
\def\paw#1#2#3{     {\it Sitzungsber Preuss. Akad. Wiss. }{\bf{#1},} (#2) #3 }
\def\grg#1#2#3{     {\it Gen. Relativ. Gravit. }{\bf{#1},} (#2) #3 }
\def\ap#1#2#3{     {\it Ann. of Phys. }{\bf{#1},} (#2) #3 }

\def\prtp#1#2#3{     {\it Prog. Theor. Phys. }{\bf{#1},} (#2) #3 }
\def\pros#1#2#3{     {\it Proc. Roy. Soc. London A }{\bf{#1},} (#2) #3 }
\def\araa#1#2#3{     {\it Annu. Rev. Astron. Astrophys. }{\bf{#1},} (#2) #3 }
\def\ijtp#1#2#3{   {\it Int. J. of Theor. Phys. } {\bf{#1},}  (#2) #3 }

\def\plb#1#2#3{    {\it Phys. Lett. }{ B \bf{#1},} (#2) #3 }
\def\pla#1#2#3{    {\it Phys. Lett. }{ A \bf{#1},} (#2) #3 }

\def\prd#1#2#3{    {\it Phys. Rev. }{ D \bf{#1},} (#2) #3 }

\def\mpla#1#2#3{    {\it Mod. Phys. Lett. }{A \bf{#1},} (#2) #3 }

\def\eq#1{{eq.~(\ref{#1})}}
\def\eqs#1#2{{eqs.~(\ref{#1})--(\ref{#2})}}
\def\abs#1{\left| #1\right|}

\def\etal{{\it et al.}}
\def\ie{{\it i.e. }}

\newcommand{\bea}{\begin{eqnarray}}
\newcommand{\beq}{\begin{equation}}
\newcommand{\eea}{\end{eqnarray}}
\newcommand{\eeq}{\end{equation}}
\newcommand{\nnu}{\nonumber}
\newcommand{\al}{\alpha}
\newcommand{\be}{\beta}
\newcommand{\ga}{\gamma}

\newcommand{\de}{\delta}
\newcommand{\si}{\sigma}

\newcommand{\la}{\lambda}
\newcommand{\La}{\Lambda}
\newcommand{\ka}{\kappa}
\newcommand{\rh}{\rho}

\newcommand{\Om}{\Omega}

\newcommand{\half}{\textstyle \frac{1}{2} \displaystyle}

\newcommand{\m}{\mu}
\newcommand{\n}{\nu}

\newcommand{\dud}[4]{{\smash{{{#1}_{#2}}^{#3}}}_{#4}}
\renewcommand{\sqrt}[1]{\left( #1 \right)^{1/2}}

\begin{document}
%%%%%%%%%%%%%%%%%%%%%%%%%
\title{\bf The Cosmology  of Tetradic Theory of Gravitation}
{\small
\author{H. A. Alhendi$^1$\ , \ E. I. Lashin$^{1,2}$ and \ G. L. Nashed$^3$\\
$^1$ Department of physics and Astronomy, College of Science,\\ King Saud University, Riyadh,
Saudi Arabia \\
$^2$ Department of Physics, Faculty of Science, \\
Ain Shams University, Cairo, Egypt\\
$^3$ Department of Matematics, Faculty of Science, \\
Ain Shams University, Cairo, Egypt\\
Email: alhendi@ksu.edu.sa, lashin@ksu.edu.sa}
}
\maketitle
\begin{abstract}
We consider a special class of the tetrad theory of gravitation
which can be considered as a viable alternative gravitational
theories. We investigate cosmological models based on those
theories by examining the possibility of fitting the recent
astronomical measurement of supernova Ia magnitude versus shift.
Our investigations result in a reasonable fit
for the supernova data without introducing a cosmological constant.
Thus, cosmological models based on tetradic theory of gravitation
can provide alternatives to dark energy models.
\\ \\
PACS numbers:\ 04.50.+h, 04.80.Cc
\end{abstract}
\section{Introduction}

The present observation of the distant supernovae type Ia
indicates that the universe is presently accelerating\cite{sp1}--\cite{res2}.
The cosmic acceleration is attributed to the presence of unknown form of
energy violating the strong energy conditions $\rh_X + 3 p_X >0$
where $\rh_X$ and $p_X$ are energy density and pressure of dark
energy respectively. Different candidates for dark energy are attempted to
yield accelerating cosmologies at late time\cite{mod1}--\cite{mod10}.
The cosmological constant $\La$ and phantom fields
\cite{phant1}--\cite{phant3} violating a weak energy conditions
$\rh_X +  p_X >0$ are most popular ones.

Recently, different attempts \cite{modg1}--\cite{modg8} have been
carried out to modify gravity to yield accelerating cosmologies at
late times. In this paper, we exploit the possibility of modifying gravity
based on absolute parallelism spaces.

The notion of absolute parallelism was first introduced in physics by
Einstein\cite{ein} trying to unify gravitation and electromagnetism
into 16 degrees of freedom of the tetrads. His attempt failed,
however, because there was no Schwarzchild solution in his field
equations.

The interest in the tetrad theory as a purely gravitational theory
was revived by M\o ller\cite{moller} who showed that a more satisfactory
treatment of energy momentum complex than that of general
relativity can be achieved. In his first attempt in finding
Lagrangians  M\o ller's  was restricted by the assumption that the equations
determining the metric tensor should coincide with the Einstein equations.
After then, he \cite{moller-78} abandoned this assumption and relooked
for a wider class of Lagrangians, allowing for possible deviation
from the Einstein equations in the case of strong gravitational
fields. M\o ller's theory was generalized into scalar tetradic
theory by S\' aez\cite{saez-83}. Meyer showed that M\o ller's theory is a
special case of the Poincare gauge theory\cite{meyer}.

Quite independently, Hayashi and Nakano\cite{hn} formulated the
tetrad theory of gravitation as a gauge theory of the space-time
translation group. Hayashi and Shirafuji\cite{hs} studied the
geometrical and observational basis of the tetrad theory, assuming
that the Lagrangian be given by a quadratic form of torsion tensor.
If the invariance under the parity operations is assumed, the most
general Lagrangian consists of three terms with three unknown
parameters to be fixed by experiment. Two of the three parameters
were determined by comparison with solar-system experiments,
while only an upper bound has been estimated for the third
\cite{hs}--\cite{mn}.

The numerical values of the two parameters found were very small,
consistent with a value of zero. If these two parameters are
exactly equal to zero, the theory reduces to the one proposed by
Hayashi and Nakano and M\o ller which we shall here refer to as
the {\bf HNM} theory. This theory differs from
general relativity only when the torsion tensor has a nonvanishing
axial vector part.

Many applications of HNM theory have been done during the years.
These include cosmological applications \cite{saez-84a},investigating
gravitational radiation \cite{gr1}--\cite{gr2} and energy momentum
complex \cite{mikhail1}--\cite{nash},
and finding a general solution with spherical symmetry\cite{mikhail2},
and solution with axial symmetry in \cite{saez-84b}.

The rest of the paper is organized as follows: In section~2 we
review HNM tetrad theory of gravitation. In section~3 we present
the tetrad field satisfying the requirement of homogeneity and
isotropy. In section~4 we discuss briefly the basic of the tetrad
cosmology and derive the relation between the luminosity distance
and redshift. In section~5 we present numerical investigations of
the tetrad cosmological model to fit the recent observational supernovae
data. The obtained results are compared with standard cosmology (with or without
cosmological constant). Finally, in the last section we give our
discussion.

\section{HNM Tetrad Theory of Gravitation}

In this paper we follow M\o ller construction\cite{moller-78}of the tetrad
theory of gravitation based on the Weitzenb\o ck space-time. In this theory
the field variables are the 16 tetrad components~${e_i}^\mu$, from
which the metric is constructed as
\beq
g^{\mu\nu} := \eta^{ij}{e_i}^\mu {e_j}^\nu,
\label{met}
\eeq
where $\eta^{ij}$ is the Lorentz metric tensor taken as
$\mbox{diag}(-1,1,1,1)$.
The Latin indices
$(i,j\ldots)$ refer to vector numbers and Greek indices $(\mu,\nu\ldots)$
to vector components, and  all of them run from $0$ to $3$. We restrict
indices $a,b,\ldots$ and $\alpha,\beta,\ldots$
(beginning of Latin and Greek alphabetic)  for spatial
components.

An invariant Lagrangian ${\mathcal L}$ is constructed from
$g^{\mu\nu}$ and $\ga_{\mu\nu\rho}$, where $\ga_{\mu\nu\rho}$ is
the contorsion tensor given by:
\beq
\ga_{\mu\nu\rho} := \eta^{ij}\,e_{i\mu}\, e_{j\nu;\rho},
\label{con}
\eeq
where the semicolon denotes covariant differentiation using the
Christoffel symbols. The most general Lagrangian density invariant under
the parity operation can be constructed as a linear combination of
the following expressions:
\beq
L^{(1)} := \Phi_\mu \Phi^\mu, \qquad
L^{(2)} := \ga_{\mu\nu\si}\ga^{\mu\nu\si}, \qquad
L^{(3)} := \ga_{\mu\nu\si}\ga^{\si\nu\mu},
\label{parts}
\eeq
where $\Phi_\mu$ is the basic vector defined by
\beq
\Phi_\mu := {\ga^\nu}_{\mu\nu}.
\label{bv}
\eeq
These expressions $L^{(i)}$ in \eq{parts} are homogeneous quadratic
functions in the first order derivatives of the tetrad field components.

M\o ller considered the simplest case, in which the Lagrangian
${\mathcal L}$ is a
linear combination of the quantities $L^{(i)}$, i.e., the Lagrangian
density is given by
\beq
{\mathcal L}_{\mbox{M\o ller}} := (-g)^{1/2}\, (\al_1 L^{(1)} + \al_2
L^{(2)} + \al_3 L^{(3)} ),
\label{lag}
\eeq
where
\beq
g := \det(g_{\m\n}).
\eeq

For this choice, the constants $\al_i$ had been chosen such that this
theory gives the same results as GR in the linear approximation of weak
fields.
According to his calculations, one can easily see that if one
chooses
\beq
\al_1 = -1, \qquad \al_2=\la, \qquad \al_3=1-2\la,
\eeq
with $\la$ equals to a free dimensionless parameter of order unity,
the theory will be in agreement with GR to the first order of
approximation. The same identification of the parameters
was obtained by Hayashi and Nakano\cite{hn}

M\o ller applied the action principle to the Lagrangian density
\eq{lag} and obtained the field equations in the form\cite{moller-78}
\bea
G_{\mu\nu}+H_{\mu\nu} &= -\ka T_{\mu\nu}, \nnu\\
F_{\mu\nu}  &= 0,
\label{feq}
\eea
where the Einstein tensor $G_{\mu\nu}$ is defined by
\beq
G_{\mu\nu} := R_{\mu\nu} - {1\over 2}\,g_{\mu\nu}\,R.
\label{ngr}
\eeq
Here $H_{\mu\nu}$ and $F_{\mu\nu}$ are given by
\beq
H_{\mu\nu} := \la \left[ \ga_{\al\be\mu}
{\ga^{\al\be}}_{\nu} + \ga_{\al\be\mu}
{\ga_\nu}^{\al\be} + \ga_{\al\be\nu}
{\ga_\mu}^{\al\be} + g_{\mu\nu} \left(
\ga_{\al\be\si}  \ga^{\si\be\al} - \half
\ga_{\al\be\si}  \ga^{\al\be\si} \right) \right]
\eeq
and
\beq
F_{\mu\nu} := \la \left[ \Phi_{\mu,\nu} - \Phi_{\nu,\mu} -
\Phi_{\al} \left( {\ga^\al}_{\mu\nu} - {\ga^\al}_{\nu\mu}
\right) + \dud\ga{\mu\nu}\al{;\al} \right].
\label{deff}
\eeq
The term $H_{\mu\nu}$ by which equations \eq{feq}
deviate from Einstein's field equations increases with
$\la$, which can be taken of order unity without destroying the
first order agreement with Einstein's theory in case of weak
fields.

M\o ller assumed that the energy-momentum tensor of matter fields
is symmetric . In the Hayashi-Nakano theory, however, the
energy-momentum tensor of spin-${1/2}$ fundamental particles has
a nonvanishing antisymetric part arising from the effects due to
the intrinsic spin.

\section{Tetrad fields for applications to cosmology}
The tetrad fields satisfying the symmetry requirements of homogeneity
and isotropy have been given by Robbertson\cite{robert}. It was found that there
are two possible teterads, which in Cartesian coordinate can be
written in the form
\bea
{e_0}^0 = 1,& \qquad {e_a}^0 =0,\qquad {e_0}^\al=0,\nnu\\
R(t)\,{e_a}^\al=& {\delta_a}^\al h_{-} +
{k\over 2}\,x^\al x^a \pm k^{1\over 2}\epsilon_{\al a \be}\; x^\be,
\label{cost1}
\eea
\bea
{e_0}^0 ={h_{-}\over h_{+}} ,
\qquad {e_a}^0 =\pm\,{(-k)^{1\over 2}\over h_{+}}\,x^a ,
\qquad R(t)\,{e_0}^\al= \pm\,(-k)^{1\over 2}\,x^\al,&\nnu\\
\;\;\;\;\;\;\;R(t)\,{e_a}^\al = {\delta_a}^\al h_{+} -
{k\over 2}\,x^\al x^a,&
\label{cost2}
\eea
where the constant $k$ takes the values $+1$,$-1$, or zero.
while $h_{\pm}$ and $r^2$ are defined by
\beq
h_{\pm}=1\pm{k\over 4}\,r^2\;\;,\;\;r^2= x^2 +y^2 +
z^2.
\label{hp}
\eeq
Here, $\epsilon_{\al a \be}= \pm 1$ when $(\al a \be)$ is an
even or odd permutation of $(123)$ and $0$ otherwise.

Both of these two tetrads  through \eq{met} lead to Roberston-Walker metric
given by
\beq
ds^2= -dt^2 + R(t)^2\,h_{+}^{-2}\,(dx^2 + dy^2 + dz^2),
\label{rob}
\eeq
Explicit calculations based on \eqs{con}{bv} and the
Christoffel symbols of the metric defined by \eq{met} result in
the following nonvanishing components of the tensor $\ga$ and $\Phi$
for the first tetrad in \eq{cost1}.
\bea
\ga^0_{\;ab}=-\,\de_{ab}\,R\,\dot{R}\,h_{+}^{-2}, &
\ga^a_{\;0b}=-\,\de_{ab}\,{\dot{R}\over R},\nnu\\
\ga^a_{\;bc}=-\,\epsilon_{abc}\; k^{{1\over
2}}\,h_{+}^{-1}, & \Phi_0= -\,3{\dot{R}\over R},\;\; \Phi_a=0,
\nnu\\
\eea
where $\dot{R}={dR\over dt}$.

Concerning the second tetrad, because the physically relevant second rank
tensors derived from it, could be complex as pointed out
in\cite{macrea}. Thus, it will be dropped out in the present work.

\section{The basic equations of cosmology}
The skew symmetric part of the field equations in \eq{feq} is
satisfied identically by the tetrad defined by \eq{cost1}, which
can be verified by explicit calculations. Assuming the energy-momentum
tensor for a perfect fluid, then the symmetric part of the field equations
 in \eq{feq} reduces to
\beq
\left(2\ddot{R}/ R\right) + \left(\dot{R}/ R\right)^2 + D/R^2 = -p,
\label{ddr}
\eeq
\beq
3\left(\dot{R}/ R\right)^2 + 3\,D/ R^2 = \rho,
\label{dr}
\eeq
where $D=k\,(1-3\,\la)$, $p$ is the pressure, and $\rho$
is the density of energy associated with the fluids.
Through out this work we employ units in
which $G=1/8\pi$ and $c=1$, and in this section we use the
notations and equations (with suitable modification) presented in
\cite{turner}

As in the standard cosmology, the Hubble parameter $H$ is defined by
$H = \dot{R}/R$. Then critical density is given as
\beq
\rho_c =3\left(\dot{R}/ R\right)^2= 3 H^2
\label{rc}
\eeq
Many useful parameters can be defined, among them
the density $\Om$ and the deceleration parameters defined
respectively as:
\beq
\Om = \rho/\rho_c\;\;,\;\; q= -\ddot{R}\,R/\dot{R}^2,
\label{cp}
\eeq
In this manner, \eq{ddr} and \eq{dr} take the form
\beq
q=\Om/2 + p/2\,H^2\;\;,\;\; \Om + \Om_D = 1,
\label{om}
\eeq
where $\Om_D = -D/R^2\,H^2$. It is worthy to mention that the
metric in \eq{rob} and the field equations \eqs{ddr}{dr}
involve two different constants, namely $k$ and $D$ respectively.
Since $\la$ is completely free, then the two constants are
independent. Thus, there is no unique relation between the
geometry of the universe and its fate, in contrast to the case
of general relativity (without cosmological constant), where such
a relation exists.

For matter dominated universe \eq{dr} can be written in the form
\beq
\dot{R}^2 - {\rho_0\,R_0^3\over R} =-D,
\label{mm}
\eeq
here $D$ can take any real value in contrast to the case of
general relativity where $k$ is restricted to $\pm 1,0$ values.

Let us define the parameters
\bea
a={1\over 1 + z} = {R\over R_0}, & \tau = H_0 t
\label{dfa}
\eea
where $z$ is the red shift. In terms of these parameters
\eq{mm} can be transformed into
\beq
\left( {da\over d\tau}\right)^2 = 1 + \Omega_0\,\left( {1\over a}
-1 \right),
\label{eva}
\eeq
which  can be converted, in terms of $z$  into
\beq
{dz\over d\tau} = \left(1 +
z\right)^2\,\left(1+\Omega_{m0}\,z\right)^{1\over 2}.
\label{dzt}
\eeq
The time lapse between the present time and the emission time of
light which suffers redshift $z_1$ can be obtained from \eq{dzt}
leading to
\beq
t_0 - t_1 = \int_0^{z1} {1\over \left(1 + z\right)^2}\,
\left( 1 + \Omega_{m0}\,z\right)^{-{1\over 2}} dz.
\label{age}
\eeq

Light signal moves along null geodesic whose equation is
\beq
{dr\over dt} = {\left(1- k\,r^2\right)^{1\over 2} \over R},
\label{nul1}
\eeq
and using \eqs{dfa}{dzt} we obtain
\beq
{dr \over \sqrt{1-k\,r^2}} = {\left(1+z\right)\over R_0}\,dt=
{\left(1 + \Omega_{m0}\,z\right)^{-{1\over 2}}\over R_0\,H_0
\left(1+z\right)}\, dz .
\label{nul2}
\eeq
Integrating \eq{nul2} for a light traveling through the universe from $r_1$
at time $t_1$ , and reaching at $r_1=0$ at the present time $t_0$
we obtain
\beq
\int_0^{r_1}\,{dr \over \sqrt{1-k\,r^2}} =
\int_0^{z_1}\,{\left(1 + \Omega_{m0}\,z\right)^{-{1\over 2}}\over R_0\,H_0
\left(1+z\right)}\, dz
\label{nul3}
\eeq
where
\beq
\int_0^{r_1}\,{dr \over \sqrt{1-k\,r^2}} =
\left\{
\begin{array}{ll}
{1\over \sqrt{\abs{k}}}\;\sinh^{-1}{\sqrt{\abs{k}}\,r_1} & k =-1 \\
{1\over \sqrt{\abs{k}}}\;\sin^{-1}{\sqrt{\abs{k}}\,r_1} & k =1 \\
r_1 & k =0
\end{array}
\right.
\eeq
From $\Om_{D0} = -D/R^2_0\,H^2_0$, we can express $\abs{k}$ as
\beq
\abs{k}= {\abs{\Om_{D0}}\, R_0^2\,H_0^2 \over \abs{b_\la}}
\label{abk}
\eeq
where $b_\la = 1 - 3\,\la$. Using \eq{nul3}, for the case $b_\la >0$
and $k=-1$, we obtain
\beq
\sqrt{{\abs{b_\la}\over \abs{\Om_{D0}}}}\,{1\over R_0\,H_0}\,
\sinh^{-1}{\left({\sqrt{\abs{\Om_{D0}}\over \abs{b_\la} }}\,
R_0\,H_0\,r_1\right)} =\int_0^{z_1}\,{\left(1 + \Omega_{m0}\,z\right)^{-{1\over 2}}\over R_0\,H_0
\left(1+z\right)}\, dz
\label{sinh1}
\eeq
then solving for $r_1$ one gets
\beq
r_1 =\sqrt{{\abs{b_\la}\over \abs{\Om_{D0}}}}\,{1\over R_0\,H_0}
\sinh{\left(\sqrt{\abs{\Om_{D0}}\over \abs{b_\la}}\,
\int_0^{z_1}\,{\left(1 + \Omega_{m0}\,z\right)^{-{1\over 2}}\over
\left(1+z\right)}\, dz\right)}
\label{sinh2}
\eeq

The luminosity distance can be shown to be
\beq
d_L = R_0^2\,r_1/R_1,
\eeq
and with the help of \eq{sinh2} we get
\beq
d_L =\sqrt{{\abs{b_\la}\over \abs{\Om_{D0}}}}\,{\left(1+z\right)\over H_0}
\sinh{\left(\sqrt{\abs{\Om_{D0}}\over \abs{b_\la}}\,
\int_0^{z}\,{\left(1 + \Omega_{m0}\,u\right)^{-{1\over 2}}\over
\left(1+u\right)}\, du\right)}
\label{sinh3}
\eeq

\section{Testing the model against the Supernovae data}
The supernovae of type Ia (SNe Ia) serve as excellent
cosmological standard candles. The apparent magnitude of a
"standard candle" is related to its luminosity distance $d_L$
through
\beq
m(z)= M + 5\,\log_{10}{\left[{d_L \over Mpc}\right]} + 25,
\label{mz1}
\eeq
where $M$ is the absolute magnitude and is assumed to be constant
for a standard candle like Sne Ia. The apparent magnitude also can be
expressed in terms of the dimensionless luminosity distance ${\cal D}_L(z)$
as
\beq
m(z)= {\cal M} + 5\,\log_{10}{{\cal D}_L(z)},
\label{mz2}
\eeq
with
\beq
{\cal D}_L(z)= {H_0\over c}\, d_L
\eeq
and
\bea
{\cal M} &=&  M +5\,\log_{10}{\left({c/H_0\over 1 Mpc}\right)} +
25\nnu\\
&=& M -5\,\log_{10}{h} + 42.38.
\eea

\begin{figure}[hbt]
\centering
\epsfxsize=10cm
\centerline{\epsfbox{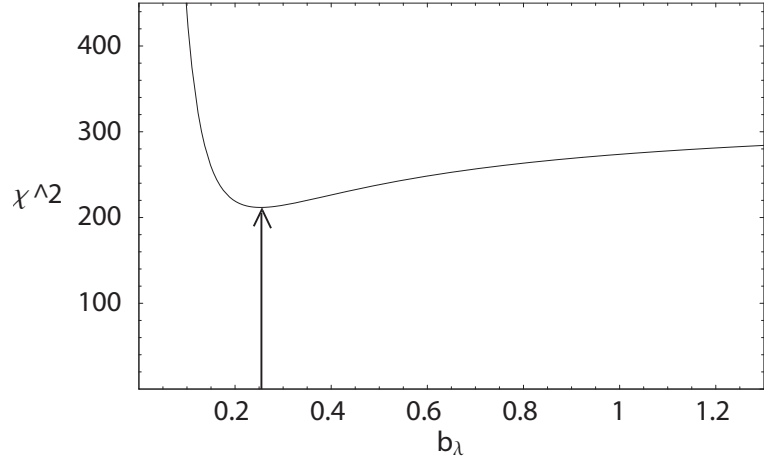}}
\caption{{\small Variation of $\chi^2$ with $b_\lambda$.}}
\label{fig1}
\end{figure}

\begin{figure}[hbt]
\centering
\epsfxsize=15cm
\centerline{\epsfbox{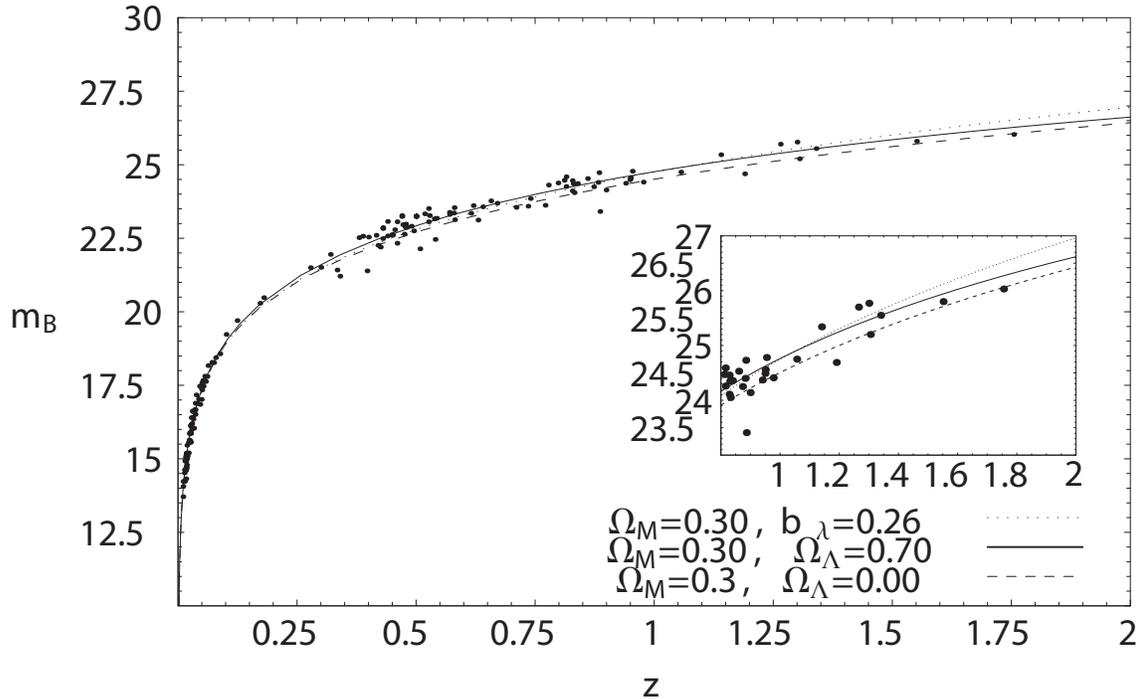}}
\caption{{\small Effective magnitude $m_{\mbox{B}}^{\mbox{eff}}$ versus
redshift $z$.}}
\label{fig2}
\end{figure}

For our present analysis we use "gold" sample compiled by Reiss \etal
\cite{res2}. The sample consists of 157 data points which in
terms of distance modulus are
\bea
\mu_{\mbox{obs}}&=&m(z) - M,\nnu \\
&=&5\,\log_{10}{{\cal D}_L(z)}-5\,\log_{10}{h} + 42.38.
\label{muo}
\eea
The best fit model to the observation is obtained by using $\chi^2$
statistics, \ie,
\beq
\chi^2 = \sum_{i=1}^{157}\,\left[{\mu_{\mbox{th}}^i - \mu_{\mbox{obs}}^i
\over \si_i}\right]^2,
\eeq
where $\mu_{\mbox{th}}$ is the predicted distance modulus for a
supernova at redshfit $z$ and $\si_i$ is the dispersion of the
mesaured distance modulus due to intrinsic and observational
uncertainties in SNe Ia peak luminosity. In our model we obtain
the best fit for $\Omega_m$ and $b_\lambda$, from the minimization
of $\chi^2$ by scanning the whole relevant parameter space, the
values we obtain are $\Omega_m=0.3$ and $b_\lambda=0.26$ for
$k=-1$ and $h=65\; \mbox{km}\; \mbox{s}^{-1} \mbox{Mpc}^{-1}$.
The other values of $k$ don't lead to a best fit. In
fig.~\ref{fig1} we
illustrate the variation of $\chi^2$ versus $b_\lambda$ for fixed
$\Omega_m=0.3$ and $k=-1$. For a one parameter fit ($b_\lambda$),
the Sne Ia  data provides the following ranges:
$0.234 \le b_\lambda \le 0.288$ at $69\%$ confidence level (CL)
and $0.207 \le b_\lambda \le 0.314$ at $95\%$ CL.
%%%%%%%%%%%%%%%%%%%%%%%%%%%%%%%%
%%%%%%%%%%%%%%%%%%%%%%%%%%%%%%%%%%%%%%%%%%%%%%%%%
%%%%%%%%%%%%%%%%%%%%%%%%%%%%%%%%%%%%%%%%%%%%%%%%%%%
\begin{table}[htbp]
\begin{center}
\begin{tabular}{cc}
\hline
\hline
$\mbox{Model}$ & $\chi^2$\\
\hline
$\Omega_M =0.30,\;\; \Omega_\Lambda = 0.70$& 178 \\
$\Omega_M =0.30,\;\; \Omega_\Lambda = 0.00$& 273 \\
$\Omega_M =0.30,\;\; b_\lambda = 0.26$& 212 \\
\hline
\hline
\end{tabular}
\end{center}
\label{tab1}
\caption{{\small Values of $\chi^2$}}
\end{table}
%%%%%%%%%%%%%%%%%%%%%%%%%%%%%%%%%%
For a quantitative comparison we use the $\chi^2$ fit. In table~\ref{tab1}
we present the $\chi^2$ values corresponding to best fit for the
general relativity with and without cosmological constant \cite{res2},
and present work. The $\chi^2$ value for
the present work seems to give a reasonable fit of the Sne Ia data without
the necessity of introducing the cosmological constant.
In Fig.~\ref{fig2}, we also make a comparison between these
theories using the effective luminosity and the redshift. Qualitatively,
as can be seen from Fig.~\ref{fig2} the tetrad theory fits the data
better than general relativity without cosmological constant and it is
almost close to the general relativity with cosmological constant for
the current available data.

%%%%%%%%%%%%%%%%%%%%%%%%%%%%%%%%%%%%%%%%%%%%%%%%%%%%%%%%%%%%%%%
\section{Discussion}
In this paper we consider a special class of the tetrad theory of
gravitation as a viable alternative gravitational theories. We
have made use of recent measurements of supernovae Ia to compare
the tetrad theory with general relativity with and without
cosmological constant. As has been shown in the previous section
the tetrad theory leads to a reasonable fit of the supernovae
data without introducing cosmological constant. However, to make a
best of the data we find that the corresponding universe is open
and non flat. This may implement a new scenario for inflationary
open universe. Such a scenario based on general relativity has been
previously discussed by \cite{linde}.

The density parameter $\Omega_M = 0.30$ of our work is the same as
that of general relativity with $\Omega_\Lambda = 0.70$, however
in our case the cosmological constant is already absent.
Concerning the age of the universe, our model gives the same value
as that of general relativity ($\Omega_M = 0.30,
\Omega_\Lambda=0$) which is $0.81 H_0^{-1}$. In the presence of
cosmological constant ($\Omega_M = 0.30,\Omega_\Lambda = 0.70$),
the age of the universe turns out to be little more around
$0.96 H_0^{-1}$.

\begin{flushleft}
{\bf Acknowledgement}
\end{flushleft}
This work was supported by Research Center at College of Science,
King Saud University under project number Phys$/1423/02$.
%\newpage
%------------------------- REFERENCES ------------------------------
\renewcommand{\baselinestretch}{1}

\end{document}